# Formation of global vortices in a dusty plasma cylinder

Zahida Ehsan[1,2,*] and Nazia Batool[3†]

[1]*Space and Plasma Astrophysics Research Group (SPAR),*

*Department of Physics, CUI, Lahore Campus 54000, Pakistan*

[2] *The Landau Laboratory for Theoretical Physics,*

*Department of Physics, CUI, Lahore Campus 54000, Pakistan.* and

[3]*National Centre for Physics, Shahdara Valley Road, Islamabad 45320, Pakistan.*[*]

(Dated: November 14, 2020)

## Abstract

The ultra low frequency modes associated to the activated dust specie in a bounded cylinder orientation plasma has been investigated here. Here flow of all plasma particles is assumed along the axial coordinates where gradients in velocity are set to be along radial direction. This analysis has been carried out for both Maxwellian and non-Maxwellian (Kappa and Carins) dusty plasma where coupling of electrostatic dust acoustic and drift modes drives global vortex formations. The coefficients of the vortex solution has been affected strongly by the presence of superthermal particles. Significance of this analysis can be viewed from the lab experiments and for the PK-4 mission perspectives. Additionally this may be fruitful for the astrophysical settings that can be mimicked as a plasma in a cylinderical column on macroscopic scale.

---

[*]Electronic address: `ehsan.zahida@gmail.com`; Electronic address: `nazia@ncp.edu.pk`

## I. BACKGROUND

Main motivation here is to investigate formation of global vortex structures in the dusty plasma of finite extent. The reason for consideration of problem is manyfold. Such as: on account of wide-ranging utilizations of dusty plasma from laboratory technology (fusion devices) to astrophysics, research in this domain has been captivating since last three decades[1–10]. The microgravity (weightlessness) environment which is now possible due to vehicles in space is ideal for experiments on dusty plasmas and for providing an excellent tool for the better understanding of physical processes like phase transitions, collective excitations because of the large time scales associated with heavier big dust particles[11–14]. Indeed dusty plasma in lab devices are of finite extent, also in principle the real plasma system contains inhomogenities in density and so can drive drift motions associated to waves. Many problems in the past have been reported where bounded plasma systems were taken into account. In the plasma system with boundaries for instance study by Vranjes and Poedts is worth mentioning where authors made a rigorous analysis of waves propagating in the inhomogenous plasma of finite extent, potential is zero in the edge region [15]. Whereas in situation of a dusty plasma wave guide, amplitudes for radial potential waves dissipates both at the center and edge of plasma column[16]. It is the reason, characteristics of wave propagations can be changed due to the geometry of the device. This is very important and indicates that propagation of waves and instabilities can be influenced by the geometry of the device where plasma system is under investigation. This clearly indicated that the most of the known theories about dispersion of complex plasma systems should be valid only under particular circumstances.

Moreover in various astrophysical situations including in the Rosette nebula column-like objects with internal filament and wave structures were indicated by the observation of the Hubble Space Telescope[17]. To study processes in these joint column shaped objects with finite boundaries, the most suitable geometry is cylindrical.

At the same time, vortices are long lived rotating structures in velocity pattern of the plasma fluid, which can be driven due to combination of flow, electric current and magnetic field. Vortices are often involved in dynamic transport. These structures can be considered as building blocks in turbulent physics. They accumulate energy in their localized patterns and carry the energy over long distances[18].

Mathematically vortex is a finite region whose vorticity is larger in amplitude than in surrounding fluid. Vortex solutions are generally in the form of Bessel functions. Vortices typically originate either by some external driving force e.g. by the spoon in the soup cup, planetary rotations and by localized heating in plasma experiments or by turbulence or self organization. Vortices are noticed in a broad range of the physical systems, ranging from quantum system to galactic systems. They are frequently observed in the ocean, in atmosphere , in earth's ionosphere and magnetosphere. The most phenomenal example of localized planetary vortex is Great Red Spot on Jupiter. This vortex is significantly larger than earth in diameter.

Vortices can be local or global and of ring shape as well, depending on the physical situation of the system. One important aspect of the interest in the study of vortical structure is that sometime growing vortex itself can be the possible free energy source and can accelerate plasma particles, one such example is formation of a ring vortex by the interaction of electromagnetic wave with dusty plasma eventually accelerating the dust particles to high velocity[19]. In addition some instabilities in plasma system may also lead to accumulation of energy at long scale length which can drive the formation of vortex like structures. Mathematically, vortices are associated with two dimensional partial differential equations, in which nonlinear term can be written in the form of Poisson bracket $[F, G] = \hat{z} \times \nabla F . \nabla G$, where $F$ and $G$ are functions of field variables. In magnetized plasmas, these nonlinearities arise from convective derivative terms ($\mathbf{v}_E \cdot \nabla$, where $\mathbf{v}_E = E \times B/B^2$ , $E$ and $B$ are usual notations for electric and magnetic fields, respectively).

For the above mentioned reasons, we believe that formation of global vortex like structures in a bounded dusty plasma will be importance in particular for the proposed experiments on dusty plasma to be carried at microgravity lab (PK4 experiment) therefore we will investigate the effects of shear flow on the nonlinear dynamics of cylindrically confined inhomogeneous dusty plasma for both Maxwellian and non-Maxwellain/non-thermal (Kappa and Cairns) distributed ions and electrons[20–23]. In support of adopting non-Maxwellain distribution function, it is worth mentioning that in one of the earlier works by Hasegawa et al.[24], author argued velocity-space diffusion obeying power law distribution such as $(\epsilon/\epsilon_0)^{-\kappa}$ is an inherent feature of any lab plasma experiment and of crucial significance in the transport properties in fact superthermal lab plasma behavior has been observed and reported[25]. Also shift of electron energy distribution function (EEDF) from Maxwellian to drifting

Maxwellian under the impact of RF electric field was emphasized in[26]. Plasma species (in particular lighter ones) which can asymptotically lead to a nonequilibrium stationary state with power-law distributions, e.g., often fitted with the Cairns, Kappa, or r, q distribution see Refs.[20–23, 27, 28]. Therefore consideration of non-Maxwellain distributed lighter species (electrons and ions in this case) will be more accurate for the formation of steady state vortex structures. To the best of the authors' knowledge, this has not been studied earlier and the results of the present investigation will be a useful addition in the existing literature of nonlinear structure in a confined dusty plasma.

## II. MATHEMATICAL FORMULATION

To study the dynamics of dust acoustic and drift waves leading to nonlinear structures in a bounded dusty plasma, its flow along the external magnetic field $B_0 = B_0\hat{\mathbf{z}}$ in a cylinder of radius $r_0$ will be considered here. Here we assume all dust particles are equally sized spherical balls of radius $a$ and in the equilibrium, whereas charge neutrality condition holds as $n_{i0} = n_{e0} + Z_d n_{d0}$ where $n_s$ is number density of $s$ species. All three components of the plasma ($s = i, e$ and negative $d$) are taken to have same sheared flow $v_{d0} = v_{i0} = v_{e0} = v_0(r)\hat{\mathbf{z}}$, therefore $B_0$ is taken to be constant. The wavelength $\lambda = 2\pi/k$ is much smaller than the gyroradius of plasma particle therefore we only consider dust grains to be magnetized while magnetization of lighter species electrons and ions is being ignored.

Generally we know dusty plasmas are low temperature systems and therefore $T_d \ll T_i, T_e$. Momentum equation for the dust under the condition of drift approximation $|\frac{\partial}{\partial t}| < \Omega_d = eB_0/m_d c$ gives

$$\mathbf{v}_d \simeq \frac{c}{B_0}\hat{\mathbf{z}} \times \nabla\phi + \frac{c}{B_0\Omega_d}\frac{d}{dt}\nabla_\perp\phi = \mathbf{v}_E + \mathbf{v}_{pd} \qquad (1)$$

where $v_E$ and $v_{pd}$ represent the electric and polarization drift of the negatively charged dust, respectively. In above equation we have used

$$\frac{d}{dt} = (\frac{\partial}{\partial t} + v_0 \cdot \nabla + v_E \cdot \nabla) \qquad (2)$$

whereas $|\mathbf{v}_{pd}|/|\mathbf{v}_E| \simeq O(\epsilon)$ where $\epsilon < 1$. Continuity equation for dust particles is given as

$$\frac{\partial n_d}{\partial t} + \nabla \cdot \{n_d \mathbf{v}_E + n_d \mathbf{v}_{pd}\} + \frac{\partial}{\partial x}(n_d v_{dz}) = 0 \qquad (3)$$

Here we will ignore the charge fluctuation effect which is an important feature of the dusty plasma because variations in the charge cause dissipation affects leading to the damping of the relevant acoustic mode[29–31]; however, some claims it is not the case[32]. Moreover to incorporate variable dust charge affect one needs to deal with the complex function of charging cross section which depends upon parameters like external magnetic field and the impact of the particles approaching grains to distances smaller than the particle size[33].

Here we assume gyroradius of electrons is much smaller than the size of dust practices and so changes in the dust charge can be negligible, in such case, electrons reach quite fast the surface of dust practices along the direction of magnetic field ( $B$) and therefore electrons move faster due to higher mobility taking place in the dust charging process.. It should be mentioned that in our analysis, the occurrence time for the formation of vortex like structure is much shorter than the time needed for further significant variation in dust charge, and therefore we do not taken into account dust charge fluctuations which is indeed an important feature of dusty plasmas.

Thus for low frequency dynamics in a magnetized plasma when $\omega/k \ll v_{ts}$ the lighter species ($m_{e,i} \to 0$) can obey Boltzmann distribution

$$n_e \simeq n_{e0}(r)\exp\left(\frac{e\phi}{T_e}\right) \simeq n_{e0}(r)\frac{e\phi}{T_e} \tag{4}$$

$$n_i \simeq n_{i0}(r)\exp\left(-\frac{e\phi}{T_i}\right) \simeq -n_{i0}(r)\frac{e\phi}{T_i} \tag{5}$$

in above background density of plasma depends upon r-coordinates. Poisson equation is given as

$$\nabla^2 \phi = 4\pi e \left(Z_d n_d + n_{e0}(r)\frac{e\phi}{T_e} - n_{i0}(r)\frac{e\phi}{T_i}\right) \tag{6}$$

Using (1-6), one gets,

$$\left(\frac{\partial}{\partial t} + v_0 \frac{\partial}{\partial z} + \frac{c}{B_0}\hat{\mathbf{z}} \times \nabla\phi \cdot \nabla\right)\left(\frac{T_i}{4\pi e^2}\nabla^2 \phi - N_0 \phi\right) + \frac{Z_d c T_i}{eB_0}\hat{\mathbf{z}} \times \nabla\phi \cdot \nabla n_{d0}$$

$$+\frac{Z_d n_{d0} c T_i}{eB_0 \Omega_d}\left(\frac{\partial}{\partial t} + v_0 \frac{\partial}{\partial z} + \frac{c}{B_0}\hat{\mathbf{z}} \times \nabla\phi \cdot \nabla\right)\nabla^2 \phi + Z_d n_{d0}\frac{\partial v_{dz}}{\partial z} = 0 \tag{7}$$

to obtain above equation, higher order terms have been ignored. Introducing normalized potential $\Phi = e\phi/T_e$, and dust gyradious $\rho_{sd} = c_{sd}/\Omega_d$ the above equation can be expressed as

$$\left(\frac{\partial}{\partial t} + v_0 \frac{\partial}{\partial z}\right)\mathbf{V}_{dz} + \frac{\Omega_d \rho_d^2}{Z_d^2 \alpha T_e}\mathbf{J}\{\Phi, \mathbf{V}_{dz}\} + \frac{\Omega_d \rho_d^2}{Z_d^2 \alpha T_e}\frac{1}{r}\frac{\partial \Phi}{\partial \theta}\frac{\partial v_0}{\partial r} = \frac{c_d^2}{Z_d^2 \alpha T_e}\frac{\partial \Phi}{\partial z} \tag{8}$$

where $c_{sd} = Z_d \left(\frac{\alpha T_e}{m_d}\right)^{\frac{1}{2}}$ is the dust acoustic speed and $\alpha = \frac{T_i n_{0d}}{n_{0e} T_i + Z_i n_{0i} T_e}$.

Eq. (8) can be written as:

$$\left(-N_D + \lambda_D^2 \nabla_\perp^2 + \rho_d^2 \nabla^2\right)\frac{\partial \Phi}{\partial t} - v_0 N_D \frac{\partial \Phi}{\partial z} + \Omega_d \rho_d^2 (\lambda_D^2 + \rho_d^2) J\{\Phi, \nabla^2 \Phi\} + \Omega_d \rho_d^2 \frac{1}{r}\frac{\partial}{\partial r}\ln n_{d0} \frac{\partial \Phi}{\partial \theta} + \frac{\partial V_{dz}}{\partial z} = 0 \quad (9)$$

where $\lambda_D = (T_i/4\pi Z_d n_{d0} e^2)^{\frac{1}{2}}$ is the Debye length and $\{\Phi, V\} = \frac{1}{r}(\partial_r \Phi \partial_\theta V - \partial_\theta \Phi \partial_r V)$ the Poisson bracket. We assume electron and ion number density such as

$$n_{e0}(r) = N_{Ae0} \exp\left(-\frac{r^2}{r_0^2}\right) \quad (10)$$

$$n_{i0}(r) = N_{Ai0} \exp\left(\frac{r^2}{r_0^2}\right) \quad (11)$$

and

$$v_0(r) = a + br^2 \quad (12)$$

where $N_{Ae,i}$ is the density of electrons and ions at axis of the cylinder $r = 0$, $a$ and $b$ are arbitrary constants.

### III. FORMATION OF VORTEX LIKE STRUCTURES

Vortices are nonlinear dispersive waves that occupy at least two dimensional nature. When the velocity of two dimensional fluid affiliated with dispersive waves gains locally larger value than the phase velocity because of nonlinear effects, then the wave shows curving of wave front which finally leads to the creation of vortex structure. Nonlinear vortex structures have been broadly studied in last few years. One particular group of solutions is explored by Meiss and Horton [34] for drift waves in a plasma corresponding to modons or dipolar vortices, described by coupled vortices of both positive and negative potential. Whereas global vortex solutions which are another class of solution have been explored by M. Yu [35] for Hasegawa Mima (HM) equation and by Nycander [36] for flute modes in inhomogenous plasma. From mathematical point of view these two type of vortex structures differ from one another and difference arises due to the boundary conditions we use.

We assume that the nonlinear structure is rotating with constant frequency $\Omega_0$ and propagates in $(\theta, z)$-coordinates making an angle $\alpha_0$ with z-axis. Let us define a moving coordinate $\eta = \theta + \alpha_0 z - \Omega_0 t$. To obtain an analytical solution, we assume $|\alpha_0 \partial_z| \ll \Omega_0 |\partial_t|$. Then

from (9) we obtain

$$\left(\frac{\partial}{\partial \eta} - \frac{\rho_d^2}{\alpha Z_d^2 T_e}\frac{\Omega_d}{\Omega_0}D_\phi\right)V_{dz} = \left(\frac{2b\Omega_d\rho_d^2 - \alpha_0 N_0 c_d^2}{Z_d^2 \alpha T_e \Omega_0}\right)\frac{\partial \Phi}{\partial \eta} \quad (13)$$

where $D_\phi = \frac{1}{r}\left(\frac{\partial \Phi}{\partial r}\frac{\partial}{\partial \eta} - \frac{\partial \Phi}{\partial \eta}\frac{\partial}{\partial r}\right)$. Integration of Eq. (13) assuming $V_{dz}$ a linear function of $\Phi$, we get

$$V_{dz} = \left\{\frac{(2b\Omega_d\rho_d^2 - \alpha_0 N_0 c_d^2)}{\alpha Z_d^2 T_e \Omega_0}\right\}\Phi \quad (14)$$

Transforming to the $\eta$ frame, (9) gives us

$$\Omega_0\left(\left(-1 + \frac{\lambda_{Di}^2 n_{i0}}{Z_d^2 n_{d0}\alpha}\frac{T_e}{T_i}\nabla^2 + \frac{\rho_d^2 \nabla_\perp^2}{\alpha^2 Z_d^3}\right)\right)\frac{\partial \Phi}{\partial \eta} = \left(\Omega_d\rho_d^2\lambda_{Di}^2\frac{n_{i0}}{Z_d^4 n_{d0}} + \frac{\Omega_{cd}\rho_d^4 T_i}{\alpha^2 Z_d^5 T_e}\right)$$

$$\times \frac{1}{r}\left(\frac{\partial \Phi}{\partial r}\frac{\partial \Phi}{\partial \eta}\nabla^2\Phi - \frac{\partial \Phi}{\partial \eta}\frac{\partial}{\partial r}\nabla^2\Phi\right) - \frac{T_e}{T_i}\alpha_0\left(\frac{2b\Omega_d\rho_d^2 - \alpha_0 c_d^2}{\Omega_0\alpha Z_d^2 T_e}\right)\frac{\partial \Phi}{\partial \eta} \quad (15)$$

Introducing $L_\phi = \left(\frac{\partial}{\partial \eta} - \frac{\Omega_d\rho_d^2}{\Omega_0 r Z_d^2}\frac{T_i}{T_e}D_\phi\right)$ as another operator in above equation gives us:

$$L_\phi\left[\nabla^2\Phi + \left\{\frac{\frac{2}{r_0^2}\frac{\Omega_d\rho_d^2 n_{d0}}{\Omega_0^2\alpha^2 Z_d^2} - \frac{T_e}{T_i^2 Z_d\alpha}\frac{\alpha_0(2b\Omega_d\rho_d^2 - c_d^2\alpha_0)}{\Omega_0^2\alpha Z_d^2 T_e} - 1}{\left(\frac{\lambda_{Di}^2 n_{i0}}{Z_d^2 n_{d0}\alpha T_i} + \frac{\rho_d^2}{\alpha^2 Z_d^3}\right)}\right\}\Phi\right] = 0 \quad (16)$$

or

$$\nabla^2 \Phi + \chi^2 \Phi = Ar^2 \quad (17)$$

where $A = \frac{B}{2H}$ and

$$\chi^2 = \frac{\frac{2}{r_0^2}\frac{\Omega_d\rho_d^2 n_{d0}}{\Omega_0^2\alpha^2 Z_d^2} - \frac{T_e}{T_i^2 Z_d\alpha}\frac{\alpha_0(2b\Omega_d\rho_d^2 - c_d^2\alpha_0)}{\Omega_0^2\alpha Z_d^2 T_e} - 1}{\left(\frac{\lambda_{Di}^2 n_{i0}}{Z_d^2 n_{d0}\alpha T_i} + \frac{\rho_d^2}{\alpha^2 Z_d^3}\right)} - B \quad (18)$$

The general solution of (17) is

$$\Phi(r,\eta) = \Phi_m J_n(\chi r)cos(n\eta) + \frac{Ar^2}{\chi^2} - \frac{4A}{\chi^4} \quad (19)$$

where $\Phi_m$ is maximum amplitude of vortex, $J_n$ the Bessel function of order $n$. Solution (19) must satisfy the boundary conditions $\Phi = 0$ at $r = r_0$, therefore, at $r = r_0$ we require $\chi = \frac{k_a}{r_0}$ where $k_a$ are zeros of Bessel function. Then using $\Phi = 0$ at $r = r_0$, we obtain global vortex solution in a cylinder as

$$\Phi(r,\eta) = \Phi_m J_n\left(\frac{k_a r}{r_0}\right)cos(n\eta) + \frac{A_0^2 r_0^2}{k_a^2}\left(r^2 - \frac{4r_0^2}{k_\alpha^2}\right) \quad (20)$$

where

$$A_0 = \frac{1}{2H}\left[\frac{\frac{2}{r_0^2}\frac{\Omega_d\rho_d^2 n_{d0}}{\Omega_0^2\alpha^2 Z_d^2} - \frac{T_e}{T_i^2 Z_d\alpha}\frac{\alpha_0(2b\Omega_d\rho_d^2 - c_d^2\alpha_0)}{\Omega_0^2\alpha Z_d^2 T_e} - 1}{\left(\frac{\lambda_{Di}^2 n_{i0}}{Z_d^2 n_{d0}\alpha T_i} + \frac{\rho_d^2}{\alpha^2 Z_d^3}\right)} - \frac{k_a^2}{r_0^2}\right] \quad (21)$$

## IV. KAPPA AND CAIRNS DISTRIBUTION

In this section, we will develop nonlinear equation admitting vortex like solutions for the non-Maxwellian dusty plasma where lighter species behave Kappa and Cairns, all other assumptions described above will remain the same. Hence the number densities for Kappa distributed electrons and ions can be represented as follows[20, 22]:

$$n_e = n_{e0}(r) \left\{ 1 - \left(\kappa - \frac{3}{2}\right)^{-1} \frac{e\phi}{k_B T_e} \right\}^{-\kappa+1/2}, \quad (21)$$

$$n_i = n_{i0}(r) \left\{ 1 + \left(\kappa - \frac{3}{2}\right)^{-1} \frac{e\phi}{k_B T_i} \right\}^{-\kappa+1/2} \quad (22)$$

where $\kappa$ is the spectral index measuring the deviation form Maxwellian distribution and for $\kappa \to \infty$ Maxwellian distribution is achieved. Poisson equation is given as

$$Z_d n_d = \frac{T_e}{4\pi e^2} \nabla^2 \phi - \left(\frac{\kappa - 1/2}{\kappa - 3/2}\right) N_0 \phi \quad (23)$$

where $N_0 = \frac{Z_i n_{0i} T_e - n_{0e} T_i}{T_i n_{0i}}$, to obtain above equation, (21 & 22) have been expanded for $\kappa \geq 3$ ($e\phi < k_B T_{e,i}$), otherwise for $\kappa < 3$, both higher and lower order terms would have been comparable so could not be ignored. From Eqs. (1-4) & (21-23), we obtain

$$\left(\frac{\partial}{\partial t} + v_0 \frac{\partial}{\partial z} + \frac{c}{B_0} \hat{\mathbf{z}} \times \nabla \phi \cdot \nabla\right) \left[\frac{T_i}{4\pi e^2} \nabla^2 \phi - \left(\frac{\kappa - 1/2}{\kappa - 3/2}\right) N_0 \phi\right] + \frac{Z_d c T_i}{e B_0} \hat{\mathbf{z}} \times \nabla \phi \cdot \nabla n_{d0}$$

$$+ \frac{Z_d n_{d0} c T_i}{e B_0 \Omega_d} \left(\frac{\partial}{\partial t} + v_0 \frac{\partial}{\partial z} + \frac{c}{B_0} \hat{\mathbf{z}} \times \nabla \phi \cdot \nabla\right) \nabla^2 \phi + Z_d n_{d0} \frac{\partial v_{dz}}{\partial z} = 0 \quad (24)$$

And in normalized potential form (24) can be expressed as

$$\left(\frac{\partial}{\partial t} + v_0 \frac{\partial}{\partial z}\right) \mathbf{V}_{dz} + \frac{\Omega_d \rho_d^2}{Z_d^2 \alpha T_e} J\{\Phi, \mathbf{V}_{dz}\} + \frac{\Omega_d \rho_d^2}{Z_d^2 \alpha T_e} \frac{1}{r} \frac{\partial \Phi}{\partial \theta} \frac{\partial v_0}{\partial r} = \left(\frac{\kappa - 1/2}{\kappa - 3/2}\right) \frac{c_d^2 N_0}{Z_d^2 \alpha T_e} \frac{\partial \Phi}{\partial z} \quad (25)$$

Above equation can be written as:

$$\left[-\left(\frac{\kappa - 1/2}{\kappa - 3/2}\right) N_D + \lambda_D^2 \nabla_\perp^2 + \rho_d^2 \nabla^2\right] \frac{\partial \Phi}{\partial t} \times$$

$$-v_0 \left(\frac{\kappa - 1/2}{\kappa - 3/2}\right) N_D \frac{\partial \Phi}{\partial z} + \Omega_d \rho_d^2 (\lambda_D^2 + \rho_d^2) J\{\Phi, \nabla^2 \Phi\} + \Omega_d \rho_d^2 \frac{1}{r} \frac{\partial}{\partial r} \ln n_{d0} \frac{\partial \Phi}{\partial \theta} + \frac{\partial V_{dz}}{\partial z} = 0 \quad (26)$$

or

$$\left[\frac{N_D - \left(\frac{\kappa-1/2}{\kappa-3/2}\right) \left\{\frac{2}{r_0^2} \frac{c_{sd}^2 n_{d0}}{\Omega_0^2 \alpha^2 Z_d^2} - \frac{c_{sd}^2 n_{d0} \beta(2b - Z_d \beta)}{\Omega_d \Omega_0^2 \alpha Z_d^2}\right\}}{\left(\frac{\kappa-1/2}{\kappa-3/2}\right) \left(\lambda_D^2 n_{0d} Z_d + \frac{n_{0d} c_{sd}^2}{\Omega_d}\right)}\right] L_\phi \Phi = L_\phi \nabla^2 \Phi \quad (27)$$

where $L_\Phi = \frac{\partial}{\partial \eta} - H' D_\Phi$ and

$$H' = -\left(\frac{\kappa - 1/2}{\kappa - 3/2}\right) \left[\frac{\lambda_D^2 c_{sd}^2}{\alpha Z_d (\Omega_d \lambda_D^2 + c_{sd}^2)} \left\{1 + \frac{c_{sd}^2}{\alpha Z_d \Omega_d^2 \lambda_D^2}\right\}\right] \tag{28}$$

or

$$\nabla_\perp^2 \Phi + \Lambda^2 \Phi = C r^2 \tag{29}$$

where $C = \frac{C_1}{2H'}$ and

$$C_1 = \frac{\frac{2}{r_0^2} \frac{c_{sd}^2 n_{d0}}{\Omega_0^2 \alpha^2 Z_d^2} - \frac{c_{sd}^2 n_{d0} \beta (2b - Z_d \beta)}{\Omega_d \Omega_0^2 \alpha Z_d^2}}{\left(\lambda_D^2 n_{0d} Z_d + \frac{n_{0d} c_{sd}^2}{\Omega_d}\right)} - \left(\frac{\kappa - 3/2}{\kappa - 1/2}\right) N_D - \Lambda^2 \tag{30}$$

$$\Phi(r, \eta) = \Phi_m J_n\left(\frac{k_a r}{r_0}\right) cos(n\eta) + \frac{A_0^{(\kappa)^2} r_0^2}{k_a^2}\left(r^2 - \frac{4r_0^2}{k_\alpha^2}\right) \tag{31}$$

where

$$A_0^{(\kappa)} = \frac{1}{2H'}\left[\frac{N_D - \left(\frac{\kappa-1/2}{\kappa-3/2}\right)\left\{\frac{2}{r_0^2}\frac{c_{sd}^2 n_{d0}}{\Omega_0^2 \alpha^2 Z_d^2} - \frac{c_{sd}^2 n_{d0} \beta(2b-Z_d\beta)}{\Omega_d \Omega_0^2 \alpha Z_d^2}\right\}}{\left(\frac{\kappa-1/2}{\kappa-3/2}\right)\left(\lambda_D^2 n_{0d} Z_d + \frac{n_{0d} c_{sd}^2}{\Omega_d}\right)} - \frac{k_a^2}{r_0^2}\right] \tag{32}$$

The general solution of (29) is Eq. (31) represents that global vortex profile is usually determined by frequency $\Omega_0$, radius $r_0$ and maximum amplitude $\Phi_m$.

Now for the third case of Cairns distributed particles following expressions will be used [20, 23]

$$n_e = n_{e0}(r)\left\{1 - \beta^\circ\left(\frac{e\phi}{k_B T_e}\right) + \beta^\circ\left(-\frac{e\phi}{k_B T_e}\right)^2\right\} \exp\left(\frac{e\phi}{k_B T_e}\right) \tag{33}$$

$$n_i = n_{i0}(r)\left\{1 + \beta^\circ\left(\frac{e\phi}{k_B T_i}\right) - \beta^\circ\left(\frac{e\phi}{k_B T_i}\right)^2\right\} \exp\left(-\frac{e\phi}{k_B T_i}\right) \tag{34}$$

where $\beta^\circ = 4\Gamma/(1+3\Gamma)$ and $\Gamma$ works out the population of nonthermal particles where Maxwellian distribution can be retrieved for $\Gamma \to 0$. After following along the same steps given in previous cases, we finally obtain

$$\Phi(r, \eta) = \Phi_m J_n\left(\frac{k_a r}{r_0}\right) cos(n\eta) + \frac{A_0^{(\Gamma)^2} r_0^2}{k_a^2}\left(r^2 - \frac{4r_0^2}{k_\alpha^2}\right) \tag{34}$$

where

$$A_0^{(\Gamma)} = \frac{1}{2H'}\left[\frac{N_D - \left(\frac{4\Gamma}{1+3\Gamma}\right)\left\{\frac{2}{r_0^2}\frac{c_{sd}^2 n_{d0}}{\Omega_0^2 \alpha^2 Z_d^2} - \frac{c_{sd}^2 n_{d0} \beta(2b-Z_d\beta)}{\Omega_d \Omega_0^2 \alpha Z_d^2}\right\}}{\left(\frac{4\Gamma}{1+3\Gamma}\right)\left(\lambda_D^2 n_{0d} Z_d + \frac{n_{0d} c_{sd}^2}{\Omega_d}\right)} - \frac{k_a^2}{r_0^2}\right] \tag{35}$$

$$H' = -\left(\frac{4\Gamma}{1+3\Gamma}\right)\left[\frac{\lambda_D^2 c_{sd}^2}{\alpha Z_d (\Omega_d \lambda_D^2 + c_{sd}^2)}\left\{1 + \frac{c_{sd}^2}{\alpha Z_d \Omega_d^2 \lambda_D^2}\right\}\right] \tag{36}$$

For the pictorial view of the vortex formation, we choose some typical dusty plasma parameters such as $m_i = 1.67 \times 10^{-27}$ $Z_i = 1, Z_d = 50$, $n_{do} = 1 cm^{-3}$, $n_{io} = 2 \times 10^8 cm^{-3}$, $T_i = 10$eV, $B = 7.5 \times 10^{-3}T$, and $T_e = 100$eV. As can be seen from the contour plots presented in Figs. (1a, 1b, 1c & 1d) for $n = 1, 2, 3, 4$, respectively, motion and size of vortex is decided by the mode number $n$, which should be smaller than the cross section of plasma column. We have also obtained vortex formation for non-Maxwellian i.e., both Kappa and Cairns distributed plasma. Also we note how increase in the value of the dust charge (no of electrons) influence the coefficient $A_0$, as can be seen in the Fig. (2), increasing the charge number reduces the strength of the coefficient $A_0$ linearly. In Figs. (2 & 3), we also note how variations in the Kappa and Gamma indices impact the coefficient $A_0$ versus dust charge number ($Z_d$), respectively. As can be noted in the green line of Fig. (2), for the fixed value of $\kappa(= 3)$, upon increasing the values of $Z_d$, strength of the $A_0$ is enhanced significantly. Also curves of coefficient reaches close to the Maxwellian (black line), when value of $\kappa$ increases. In Fig. (3) similar behaviors for the Cairns distributed ions and electrons has been observed, increase in the parameter $\Gamma(= 0.07, 0.3, 0.5)$, impacts strength of $A_0$ eventually heading towards Maxwellian curve for $\Gamma = 0.07$.

The present analysis can be useful to for the lab experiments on dusty plasma and proposed experiments to be carried at minute gravity conditions (PK-4 mission and so on) to study dispersion characteristic of the dusty plasmas. This study may also be fruitful for the astrophysical scenarios where vortex structures can be formed.

**Acknowledgement:** Useful discussions with Prof. Hamid Saleem are gratefully acknowledged.

therein.

**Figure Captions**

Fig. 1(a, b, c & d): Profile of vortices in the dusty plasma for $n = 1, 2, 3, 4$.

Fig. (2): Coefficient $A_0$ versus $Z_d$ for a different values of spectral indices $\kappa \; (= 2, 3, 5, 10)$.

Fig. (3): Coefficient $B_0$ versus $Z_d$ for a different values of non-thermal particles i.e., $\Gamma = (0.07, 0.3, 0.5)$.

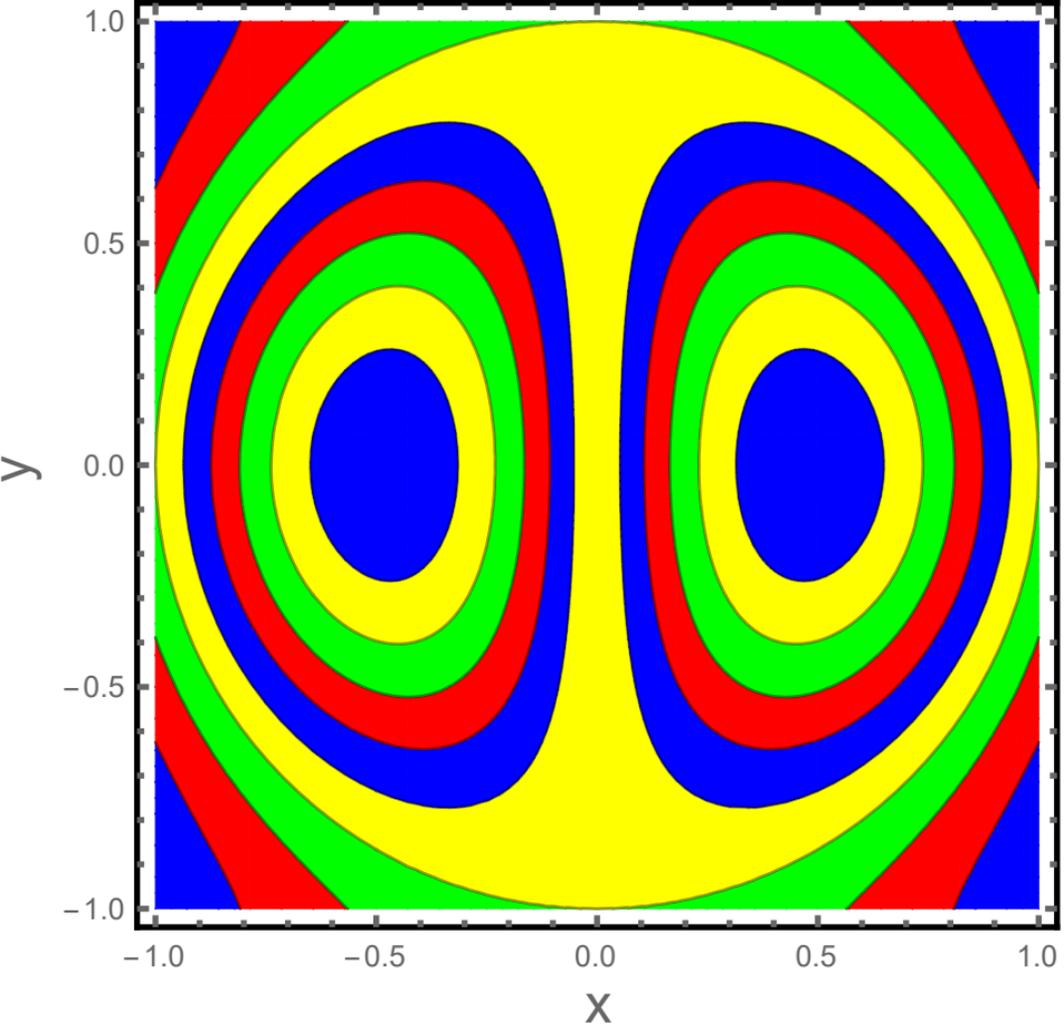

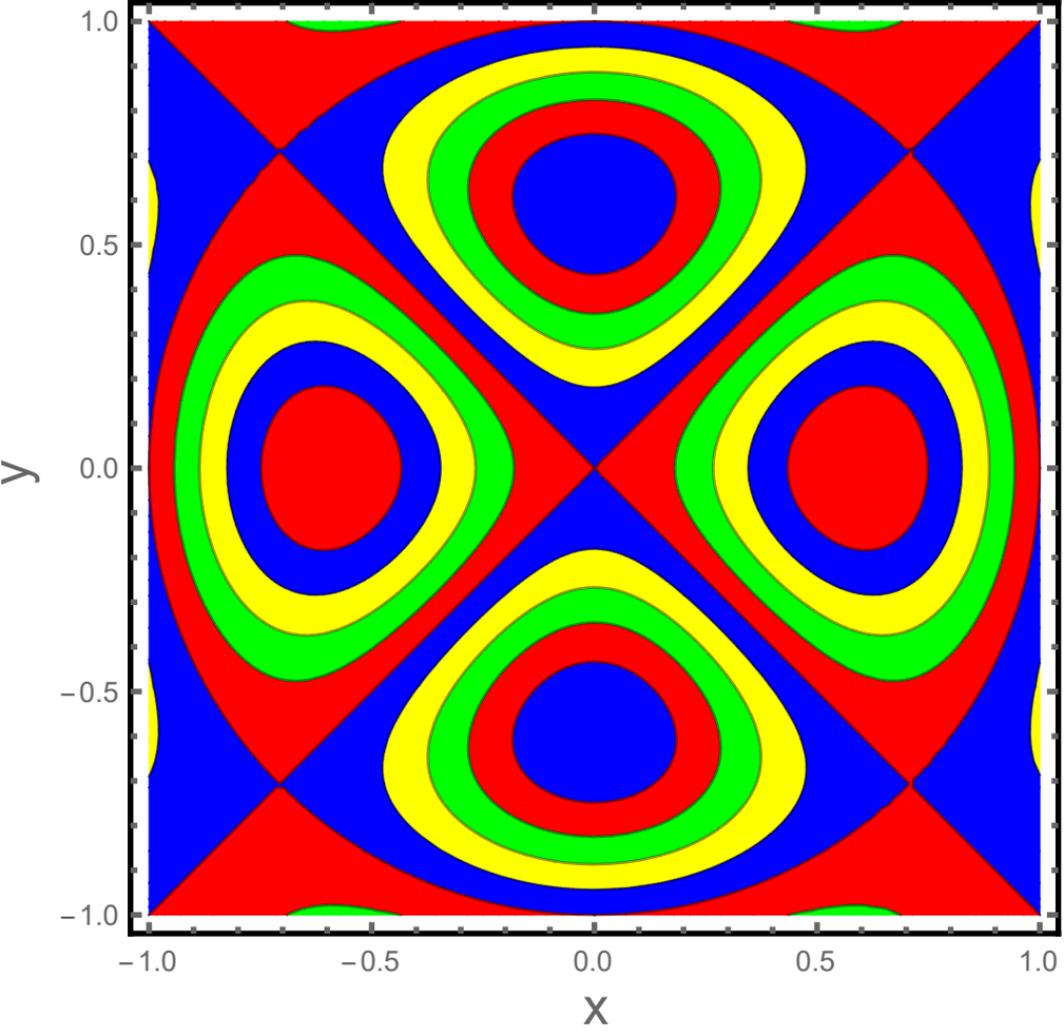

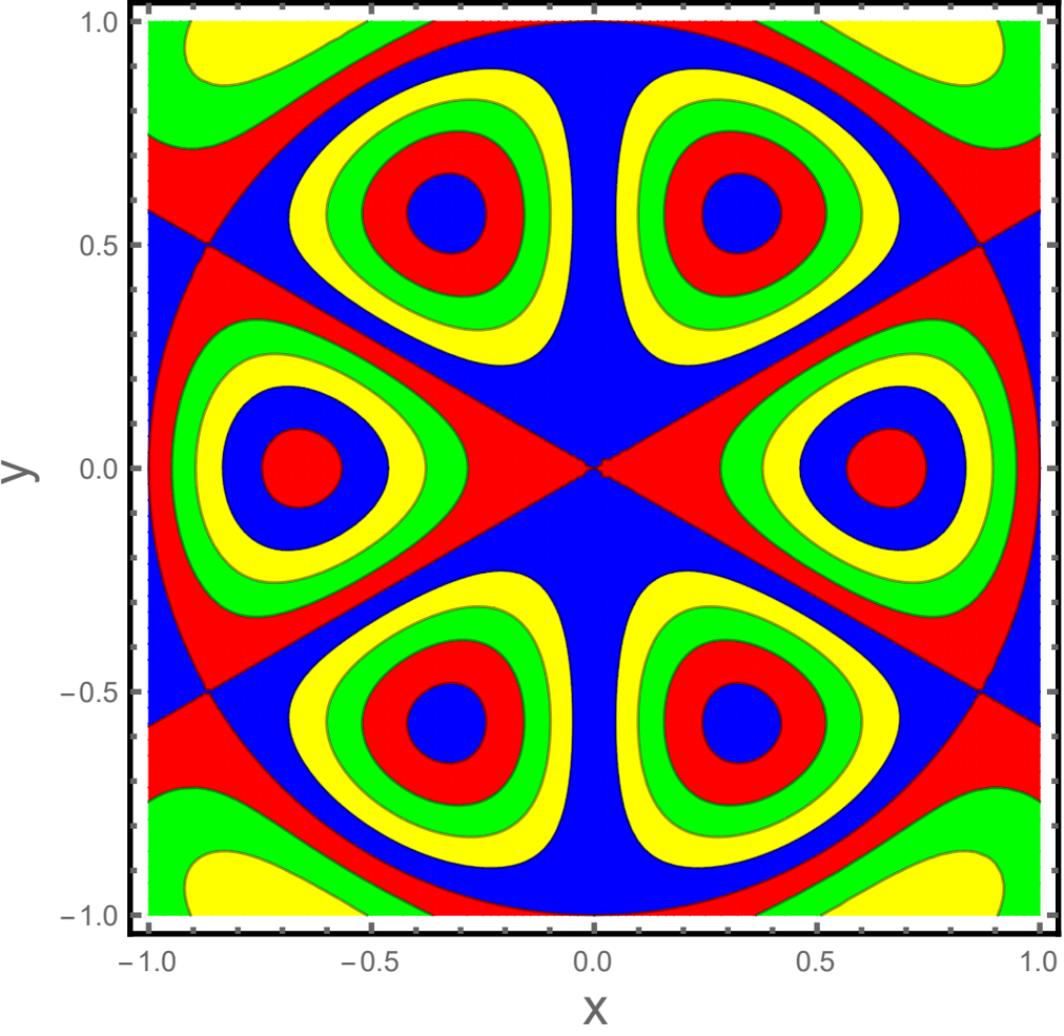

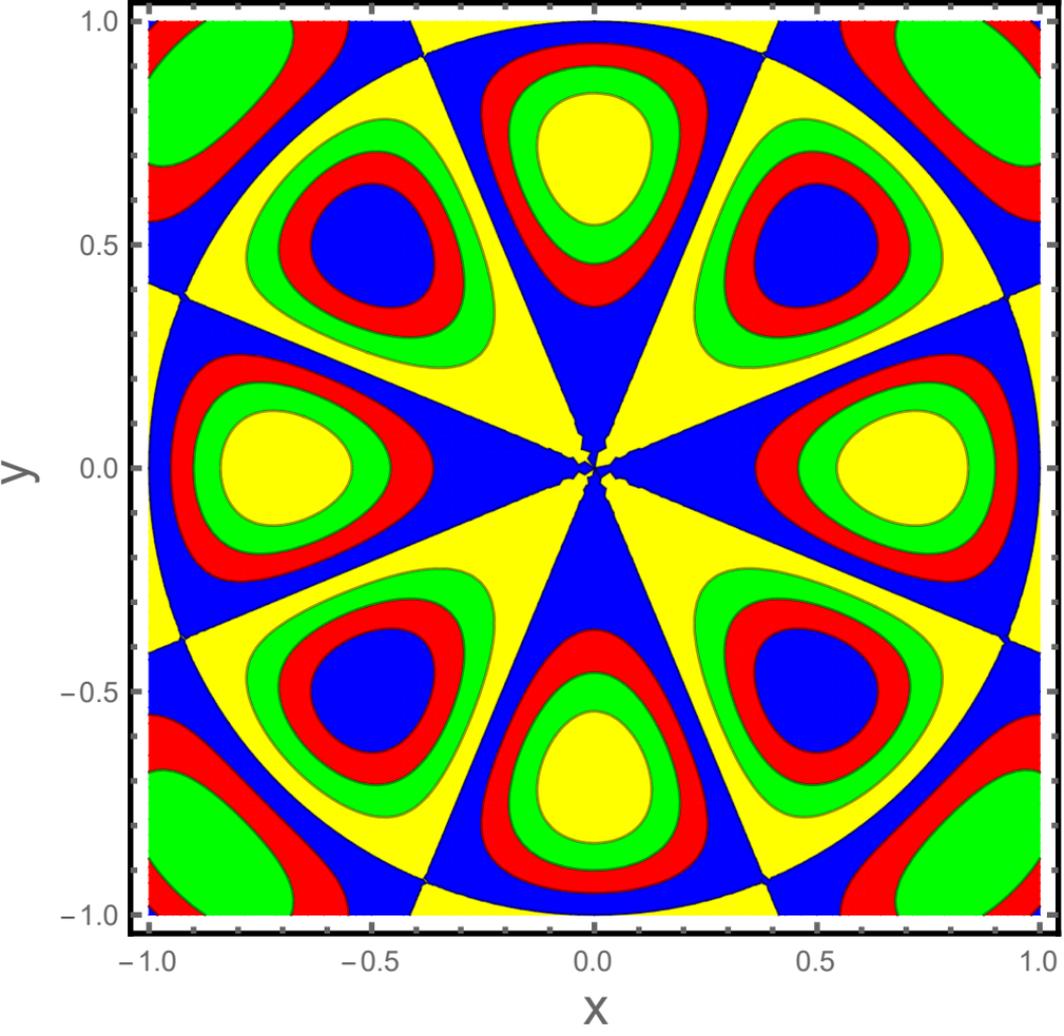

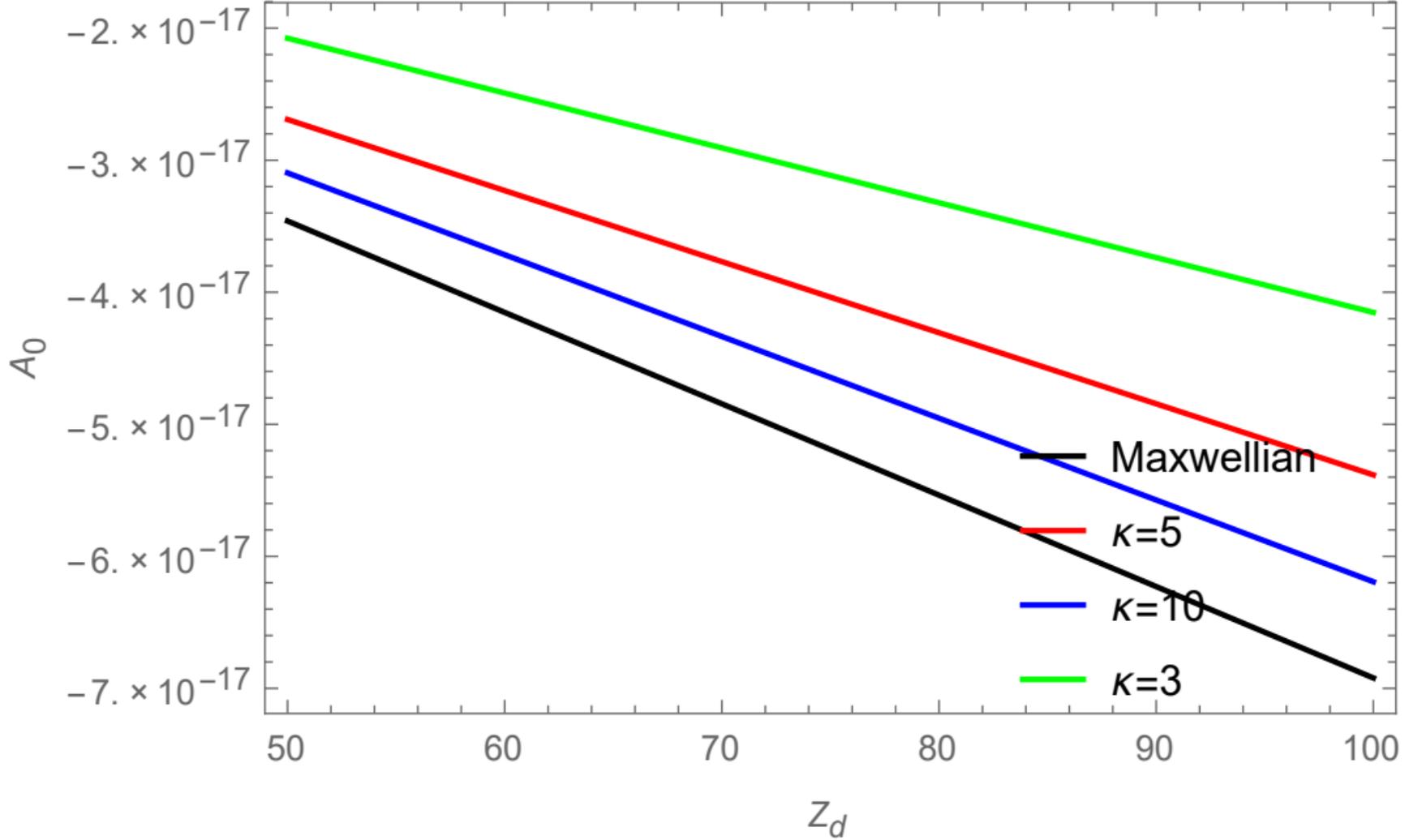

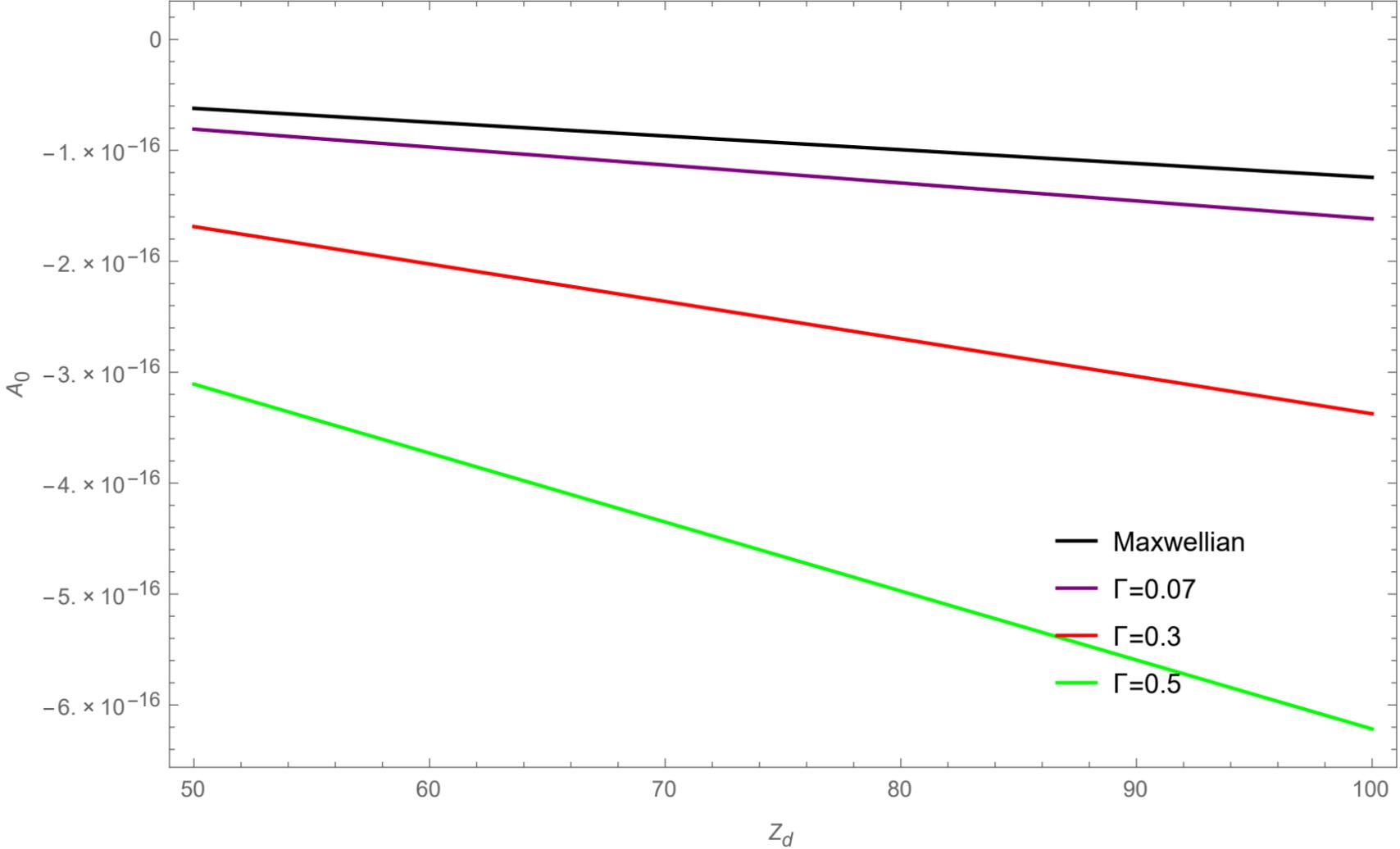